\documentclass[a4paper,10pt]{article}
\usepackage[text={7in,9in},centering]{geometry}
\usepackage[utf8]{inputenc}
\usepackage{mathtools}

\usepackage{graphicx,bm,amssymb,amsmath}
\usepackage{ulem} 

\newcommand{\brho} {\mbox {\boldmath $\rho$}}

\newcommand{\btheta} {\mbox{\boldmath $\theta$}}

\newcommand{\bj}{{\bf j}}

\newcommand{\bq}{{\bf q}}
\newcommand{\br}{{\bf r}}

\newcommand{\bz}{{\bf z}}

\newcommand{\bT}{{\bf T}}

\newcommand{\cE}{{\cal E}}

\newcommand{\cI}{{\cal I}}

\newcommand{\cR}{{\cal R}}

\newcommand{\cT}{{\cal T}}

\newcommand{\D}{\mathbb{D}}

\newcommand{\R}{\mathbb{R}}

\def\tensor#1{\protect\@ontopof{#1}{\leftrightarrow}{1.15}\mathord{\box2}}

\title{The eigenvalues and  eigenfunctions of the toroidal dipole  operator in a mesoscopic system}
\author{
	Drago\c s-Victor Anghel\thanks{Institutul National de Cercetare-Dezvoltare pentru Fizica si Inginerie Nucleara Horia Hulubei, dragos@theory.nipne.ro; \textit{corresponding author}}, \ and \
	Mircea Dolineanu\thanks{Institutul National de Cercetare-Dezvoltare pentru Fizica si Inginerie Nucleara Horia Hulubei, University of Bucharest, Faculty of Physics, mircea.dolineanu@theory.nipne.ro},
}

\begin{document}

\maketitle

\begin{abstract}
We give analytical expressions for the eigenvalues and generalized eigenfunctions of $\hat{T}_3$, the $z$-axis projection of the toroidal dipole operator, in a system consisting of a particle confined in a thin film bent into a torus shape.
We find the quantization rules for the eigenvalues, which are the essential for describing measurements of $\hat{T}_3$.
The eigenfunctions are not square integrable, so they do not belong to the Hilbert space of wave functions, but they can be interpreted in the formalism of rigged Hilbert spaces as kernels of distributions.
While these kernels appear to be problematic at first glance due to singularities, they can actually be used in practical computations. In order to illustrate this, we prescribe their action explicitly and we also provide a normalization procedure.

\end{abstract}

\section{Introduction} \label{sec_intro}

The Zeldovich's prediction of analpole, a property related to the nuclei that undergo
$\beta$-decays~\cite{SovPhysJETP.6.1184.1958.Zeldovich}, eventually led Dubovik and
coworkers to the discovery of a whole new class of toroidal multipoles in
electrodynamics~\cite{Sov.24.1965.Dubovik, SovJ.5.318.1974.Dubovik}.
Since then, the physical significance of the toroidal multipoles have been highlighted
in practically all areas of physics, from particle~\cite{PhysLettB.722.341.2013.Ho, NuclPhysB.907.1.2016.Cabral, PhysRevD.32.1266.Radescu,ModPhysA.13.5257.1998.Dubovik},
nuclear~\cite{Science.275.1759.1997.Wood, AIPConfProc.477.14.1999.Flambaum}, and
atomic physics~\cite{AnnPhys.209.13.1991.Costescu}, to condensed matter
systems~\cite{PhysRev.70.965.1946.Kittel, Physics.2.20.2009.Khomskii, PU.55.557.2012.Pyatakov, JExpTPL.52.161.1990.Tolstoi, NewJPhys.9.95.2007.Fedotov, PhysRevB.84.094421.2011.Toledano, PhysRevB.101.2020.Shimada, JexpThPhys.87.146.1998.Popov, NatNano.14.141.2019.Lehmann}
and metamaterials~\cite{Nanotechnology.30.2019.Yang, Comm.Pres.2.10.2019.Savinov, Nanophotonics.7.2017.Talebi, LasPhotRev.13.1800266.2019.Gurvitz, ScRep.5.2016.Zagoskin, nanz:book.2016toroidal}.
The toroidal dipole is the lowest order term in this class and a quantum operator
$\hat{\bT}$ of components $\hat{T}_i$, $i = 1,2,3$, has been associated with
it~\cite{AnnPhys.209.13.1991.Costescu, JPA30.3515.1997.Anghel, JPhysConfSer.2090.012151.2021.Anghel, PhysicaA.2021.Dolineanu}.
When defined in the whole $\R^3$ space, the operators $\hat{T}_i$ are hypermaximal
(i.e., they have multiple self-adjoint extensions)~\cite{JPA30.3515.1997.Anghel}.
In finite systems, the operators $\hat{T}_i$ may be split into components along local
curvilinear systems of coordinates.
In Ref.~\cite{PhysicaA.2021.Dolineanu} it was shown that, if some of these coordinates are cyclic, then the components along them are self-adjoint, corresponding, therefore, to quantum observables.

\begin{figure}[t]
	\centering
	\includegraphics[height=8 cm]{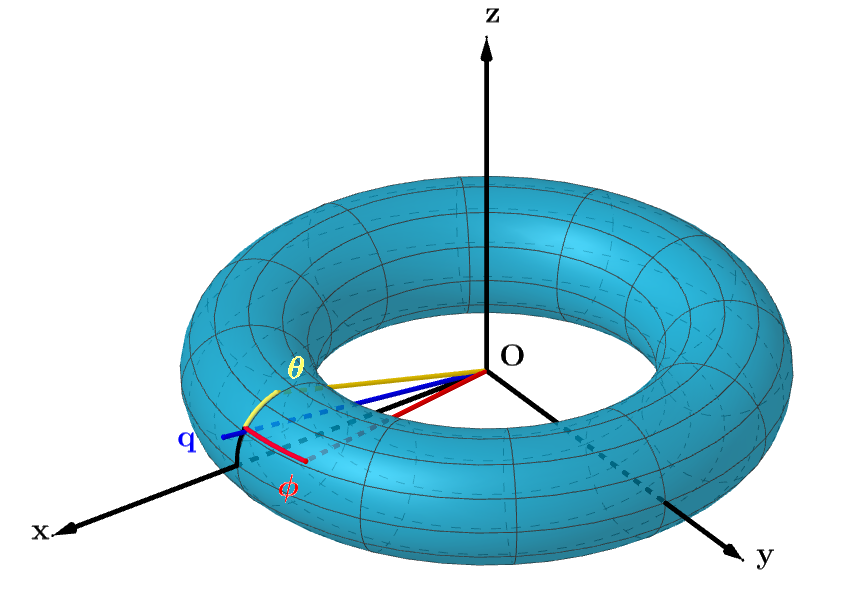}
	\caption{The system of curvilinear coordinates ($\theta$,$\phi$,$q$) on a torus
	made of a thin layer of material. The inner and outer surfaces of the material
	are two tori with the same major radius $R$, but minor radii $r$ and
	$r+q_{max}$, respectively, where $r+q_{max}<R$ and $q_{max}<<r$; that is,
	$q=0$ on the interior surface and $q=q_{max}$ on the exterior surface.}
	\label{fig_tor}
\end{figure}

In this paper we analyze the operator $\hat{T}_3$ on a torus made of a thin layer of
material, with major and minor radii $R$ and $r$, respectively, as indicated in
Fig.~\ref{fig_tor} (see also Ref.~\cite{PhysicaA.2021.Dolineanu}).
In such a system, the most convenient curvilinear coordinates are $(\theta, q, \phi)$,
defined in the figure.
Then, the component of $\hat{T}_3$ along the coordinate $\phi$ is zero, whereas
the component along $q$ is hermitian, but not self-adjoint, due to the Dirichlet
boundary conditions on the surfaces of the layer.
The component of $\hat{T}_3$ along the coordinate $\theta$, denoted here by $\hat{T}_3^{(\theta)}$, is self-adjoint when defined on the Sobolev space $H^1_{cycl}([0,2\pi])$ 
due to the periodicity of the wavefunctions~\cite{PhysicaA.2021.Dolineanu}.
While in Ref.~\cite{PhysicaA.2021.Dolineanu} the system was analyzed using a basis
formed of the eigenfunctions of the projections of the momentum operator, here we
solve directly the eigenvalues-eigenfunctions equation for $\hat{T}_3^{(\theta)}$.
We find analytical expressions for the eigenvalues and eigenfunctions.
From these, we deduce quantization relations for the eigenvalues.
The eigenfunctions are not square integrable functions, so they do not belong to the Hilbert space $H^1_{cycl}([0,2\pi])$ 
of wavefunctions and therefore must be interpreted as distributions.

The paper is organized as follows.
In Section~\ref{sec_eigfs} we find the eigenfunctions and eigenvalues of
$\hat{T}_3^{(\theta)}$, from which we deduce the quantization conditions for the
eigenvalues.
In Section~\ref{sec_projections}, we introduce the spaces of distributions
which include the eigenfunctionals of $\hat{T}_3^{(\theta)}$, namely the space of
antilinear functionals over $H^1_{cycl}([0,2\pi])$ and the dual space of
$H^1_{cycl}([0,2\pi])$, and we show that the action of these distributions on the wavefunctions in $H^1_{cycl}([0,2\pi])$ lead integrable expressions. We also provide a normalization procedure and we prescribe how wavefunctions are moved to the $\hat{T}_3^{(\theta)}$ representation.

\section{The eigenvalues and eigenfunctions of $\hat{T}_3^{(\theta)}$} \label{sec_eigfs}

The toroidal dipole of a current distribution $\bj(\br)$ is defined
as~\cite{PhysRep.187.145.1990.Dubovik}
\begin{equation} \label{def_Ti}
	\bT = \frac{1}{10} \int_V \Big[\br (\br \cdot \bj) - 2 r^2 \bj \Big] d^3\br ,
\end{equation}
where $r \equiv |\br|$ and $V$ is the volume of the system.
To the toroidal dipole one may associate an operator
$\hat{\bT}$, of components~\cite{AnnPhys.209.13.1991.Costescu, JPA30.3515.1997.Anghel}
\begin{equation} \label{def_Ti_op}
	\hat T_i \equiv \frac{1}{10 m_p} \sum_{j=1}^{3} \left(x_i x_j - 2 r^2 \delta_{ij} \right) \hat p_j
\end{equation}
in the Cartesian system of coordinates $(x_1,x_2,x_3) \equiv (x,y,z)$.

Let's analyze the toroidal dipole of a particle confined in a thin layer, bent in
the form of a torus, as shown in Fig.~\ref{fig_tor}.
The torus has the major and minor radii $R$ and $r$, respectively, where
$a \equiv R/r > 1$.
As in Ref.~\cite{PhysicaA.2021.Dolineanu}, we introduce the orthonormal curvilinear
system of coordinates $(\theta,q,\phi)$ shown in Fig.~\ref{fig_tor} and write
$\hat{T}_3$~(\ref{def_Ti_op}) in these coordinates:
\begin{eqnarray}
	\hat T_3
	&=& \frac{-i\hbar}{10 m_p} \left[
	z\rho \,\hat{\brho} - \left( 2\rho^2 + z^2 \right) \,\hat{\bz} \right] \cdot \left(
	\frac{\hat{\btheta}}{r+q} \frac{\partial}{\partial \theta}
	+ \hat{\bq}\frac{\partial}{\partial q}
	\right) ,
	\label{T_3tot}
\end{eqnarray}
where $(\rho, z, \phi)$ are the cylindrical coordinates, whereas
$\hat{\brho}, \hat{\bz}, \hat{\btheta}, \hat{\bq}$ are the unit vectors along
the indicated directions.
The expression~(\ref{T_3tot}) may further be split into two hermitian
components~\cite{PhysicaA.2021.Dolineanu},
\begin{equation}
	\hat{T}^{(q)}_3 \equiv - \frac{i\hbar}{10 m_p} \left\{\left[ z\rho \hat{\brho} - \left( 2\rho^2 + z^2 \right) \hat{\bz} \right] \cdot \hat{\bq} \frac{\partial}{\partial q}
	+ \frac{T_q(\theta,q)}{2 \rho (r+q)} \right\}
	\label{def_T3q}
\end{equation}
and
\begin{eqnarray}
	\hat{T}_3^{(\theta)} &=& - \frac {i\hbar r}{10 m_p} \left\{ - \left[ \left( 3 \cos^2\theta+1 \right) a + 2 \cos\theta \left( {a}^{2}+1 \right)  \right] \frac {\rm d}{{\rm d}\theta}
	+\frac{ \left( 9 a \cos^2\theta + 10 {a}^{2} \cos\theta + 2{a}^{3}+4\cos\theta +3a \right) \sin\theta }{2(\cos\theta + a)}
	\right\} \nonumber \\
	&=& - i C_0 \left\{ C_1(\theta,a) \frac {\rm d}{{\rm d}\theta}
	+ C_2(\theta,a) \right\} \label{def_T3l}
\end{eqnarray}
where
\begin{eqnarray}
	&& C_0 \equiv \frac {\hbar r}{10 m_p} , \quad
	C_1(\theta,a) \equiv - \left[ \left( 3 \cos^2\theta+1 \right) a + 2 \cos\theta \left( {a}^{2}+1 \right)  \right] , \nonumber \\
	&& C_2(\theta,a) \equiv \frac{ \left( 9 a \cos^2\theta + 10 {a}^{2} \cos\theta + 2{a}^{3}+4\cos\theta +3a \right) \sin\theta }{2(\cos\theta + a)} ,
	\label{defs_Cs}
\end{eqnarray}
and $T_q(\theta,q)$ is a function of $\theta$ and $q$, given in Ref.~\cite{PhysicaA.2021.Dolineanu}, but not necessary in the following calculations.

\begin{figure}
	\centering
	\includegraphics[height=6 cm]{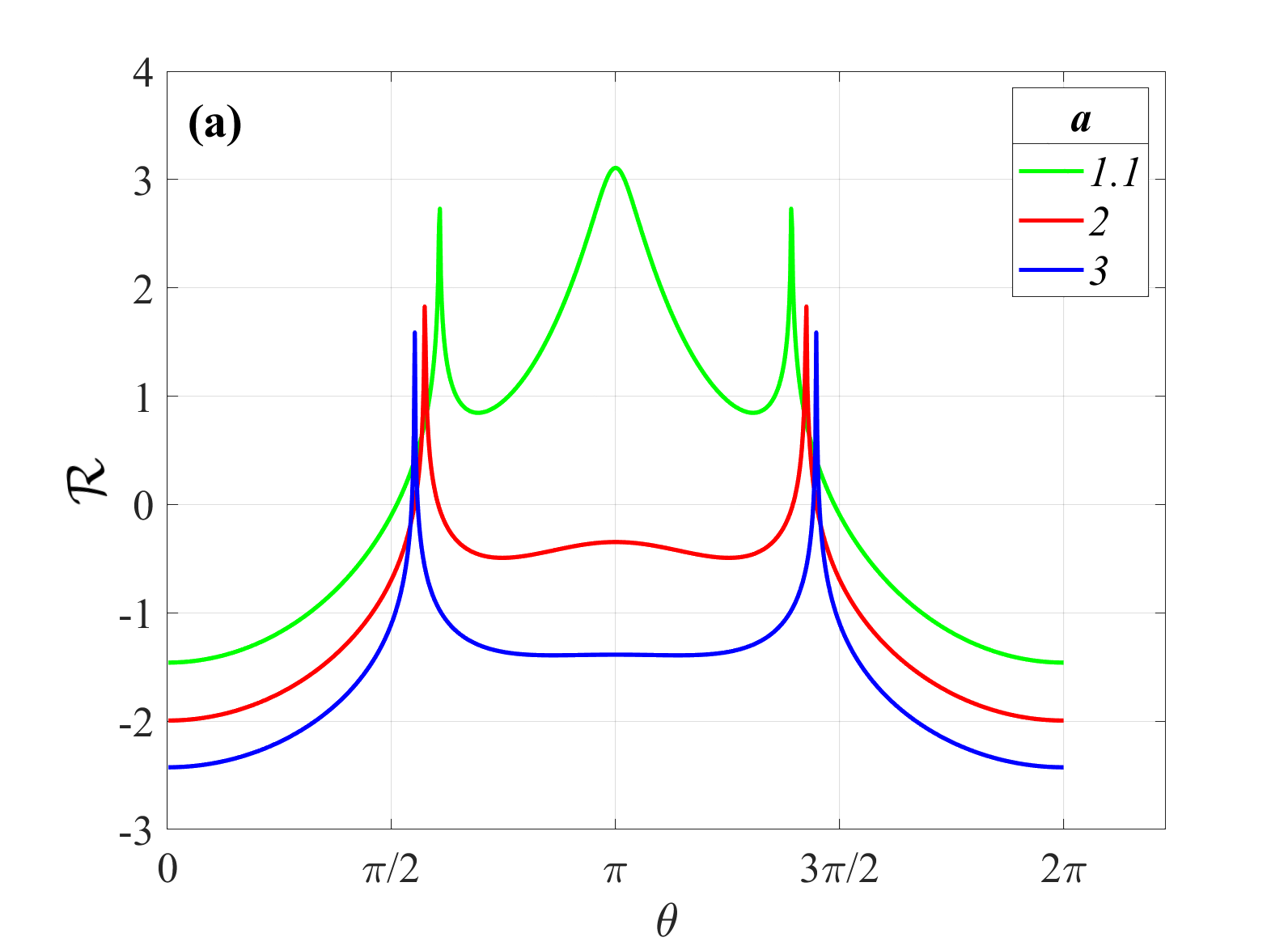}
	\includegraphics[height=6 cm]{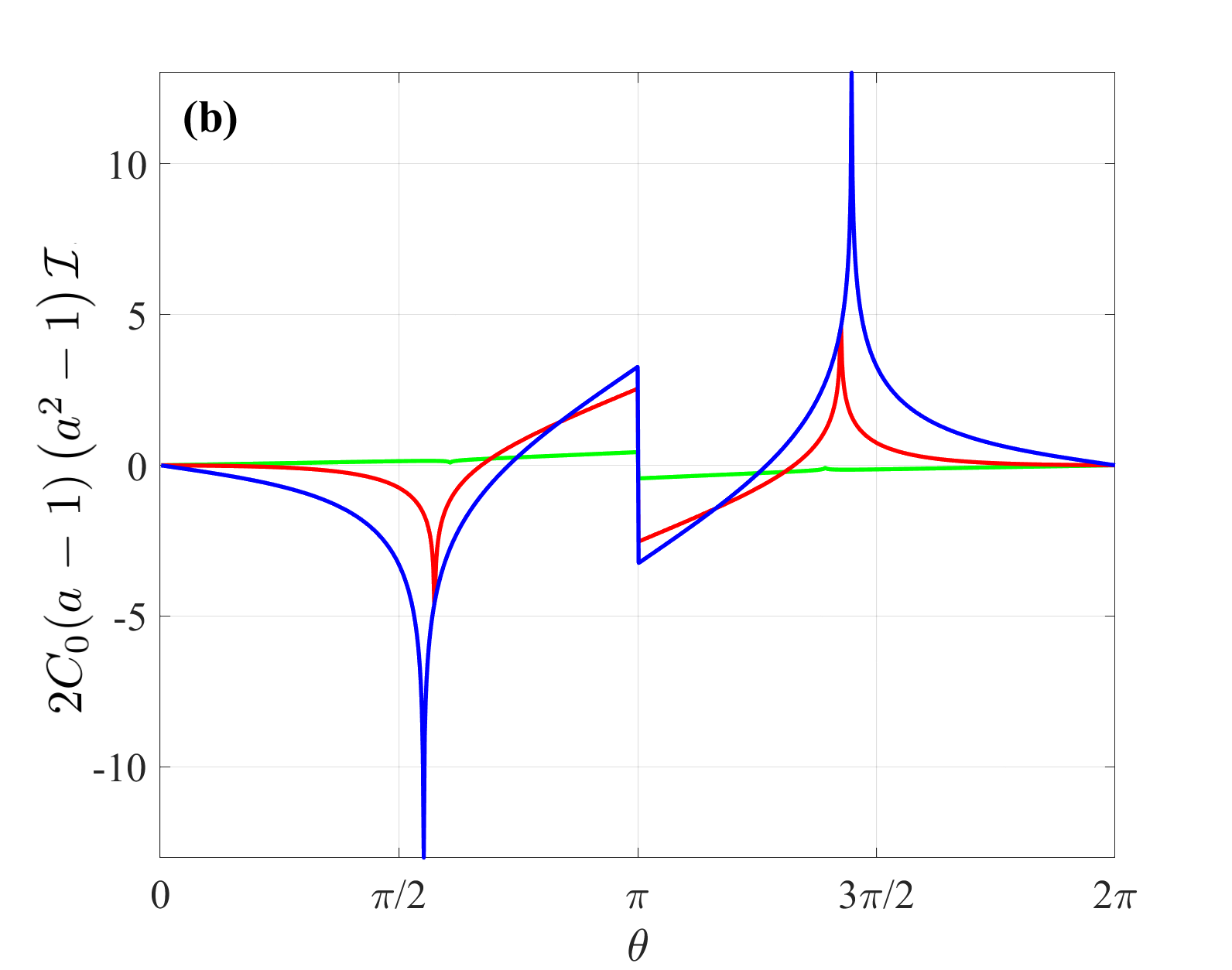}
	\caption{The real (a) and imaginary (b) parts of the primitive of $\ln\Phi$, namely $\cR$ and $2 C_0 (a-1) \left( {a}^{2}-1 \right)\cI$, respectively.
	}
	\label{fig_Imprimitive}
\end{figure}

The eigenvalues and eigenfunctions equation
\begin{subequations} \label{eigvf_T3}
\begin{equation}
	\hat{T}_3^{(\theta)} \Phi = t_3 \Phi
	\label{eigvf_T31}
\end{equation}
may be transformed into
\begin{equation}
\frac {d \Phi}{\Phi} = \frac{i \frac{t_3}{C_0} - C_2(\theta,a)}{C_1(\theta,a)} d\theta
\label{eigvf_T3l}
\end{equation}
\end{subequations}
From~(\ref{eigvf_T3l}) we obtain a relation between the primitives,
\begin{eqnarray}
	\ln(\Phi) &=&
	\frac{i}{C_0}
	\frac{t_3}{ 2 (a-1) \left( {a}^{2}-1 \right) }
	\left[
	\frac{\sqrt{\sqrt{{a}^{4}-{a}^{2}+1}+a} \left( {a}^{2}-3a+1+\sqrt {{a}^{4}-{a}^{2}+1} \right)}{2}
	\ln \left| \frac{{ {\frac{(a-1)\tan \left( {\frac {\theta}{2}} \right)}{\sqrt {\sqrt {{a}^{4}-{a}^{2}+1}-a}}}}-1}{ { \frac{(a-1)\tan \left( {\frac {\theta}{2}}\right)}{\sqrt {\sqrt {{a}^{4}-{a}^{2}+1}-a}}+1}}\right|
	\right. \nonumber \\
	&& \left. - \arctan \left( \frac {(a-1)\tan \left( {\frac {\theta}{2}} \right)}{\sqrt {\sqrt {{a}^{4}-{a}^{2}+1}+a}} \right) \sqrt {\sqrt
	{{a}^{4}-{a}^{2}+1}-a} \left( {a}^{2}-3\,a+1-\sqrt {{a}^{4}-{a}^{2}+1}
	\right)  \right] {\frac {1}{\sqrt {{a}^{4}-{a}^{2}+1}}}
	\nonumber \\
	&& - \frac {\ln  \left\{  \left( \cos\theta +a \right)
	\left| (3 \cos^2\theta+1) a + 2 \cos\theta ({a}^{2}+1) \right|  \right\} }{2}
	\equiv i t_3 \cI(\theta,a) + \cR(\theta,a),
	\label{primitive_sol}
\end{eqnarray}
where we introduced the notations $\cR(\theta,a)$ and $\cI(\theta,a)$ for the real
and imaginary parts, which are plotted in  Fig.~\ref{fig_Imprimitive}.

First, we study the divergences and the discontinuities of $\cI(\theta,a)$ and
$\cR(\theta,a)$.
We notice that $C_1(\theta,a)$, as a function of $\theta$, has zeros when
\begin{subequations}\label{eq_theta0_sols}
\begin{equation}
	(3 \cos^2\theta_0+1) a + 2 \cos\theta_0 ({a}^{2}+1) = 0 ,
	\label{eq_theta0}
\end{equation}
with the solution
\begin{equation}
\cos\theta_0 = \frac{\sqrt{a^4 - a^2 + 1} - a^2 - 1}{3a} < 0 .
\label{eq_sol_cos}
\end{equation}
Equation~(\ref{eq_sol_cos}) gives two solutions,
\begin{equation}
	\theta_0^{(1)} = {\rm arccos} \left(\frac{\sqrt{a^4 - a^2 + 1} - a^2 - 1}{3a}\right)
	\quad {\rm and} \quad
	\theta_0^{(2)} = 2\pi - {\rm arccos} \left(\frac{\sqrt{a^4 - a^2 + 1} - a^2 - 1}{3a}\right) .
	\label{eq_sols}
\end{equation}
\end{subequations}
Since $C_2(\theta_0,a) \ne 0$, then, both $\cI(\theta,a)$ and $\cR(\theta,a)$
should have logarithmic divergences at $\theta = \theta_0^{(1,2)}$.
Obviously,
\begin{subequations} \label{log_divs_P1_P2}
\begin{equation}
	\cR(\theta,a) \equiv \frac{\ln \left( \cos\theta +a \right) + \ln \left| C_1(\theta,a) \right|}{2} ,
	\label{log_div_P2}
\end{equation}
is divergent at $\theta_0$, whereas $\cI(\theta,a)$ has logarithmic divergences at
\begin{equation}
	{\frac{(a-1) \left| \tan \left( {\frac {\theta}{2}} \right) \right|}{\sqrt {\sqrt {{a}^{4}-{a}^{2}+1}-a}}}
	= 1.
	\label{log_div_P1}
\end{equation}
\end{subequations}
Replacing in~(\ref{log_div_P1}) $\theta$ by $\theta_0$ from~(\ref{eq_sol_cos}), we
obtain an identity which confirms that the divergencies appear at the same values
of $\theta$ in both, the real and imaginary parts.
For all the other values of $\theta$, both $\cI$ and $\cR$ are finite (see Fig.~\ref{fig_Imprimitive}).

\begin{figure}
	\centering
	\includegraphics[width=8 cm]{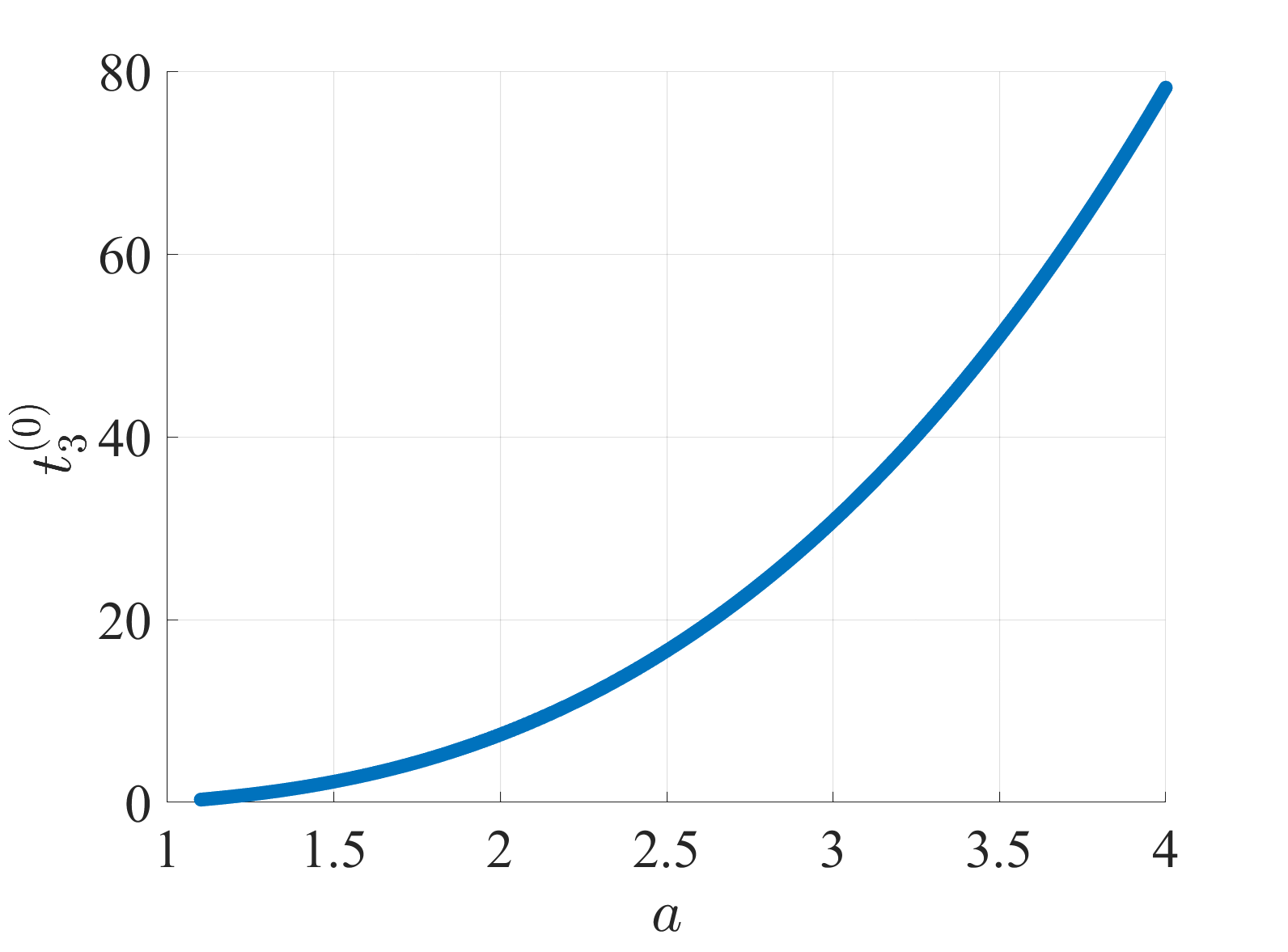}
	\caption{
	The normalized eigenvalue $t_3^{(0)}$, as a function of $a$.
	}
	\label{fig_eval}
\end{figure}

We observe that $\cR(0,a) = \cR(2\pi,a)$ and $\cI(0,a) = \cI(2\pi,a) = 0$, so the primitives are periodic.
But while $\cR(\theta,a)$ is continuous at any $\theta \ne \theta_0^{(1,2)}$, $\cI(\theta,a)$ has a jump at $\theta = \pi$ that we need to address.
From~(\ref{primitive_sol}) we obtain
\begin{equation}
	t_3 [\cI(\pi+0,a) - \cI(\pi-0,a)] = \frac{\pi t_3}{C_0}
	\frac{ \left( {a}^{2}-3\,a+1-\sqrt {{a}^{4}-{a}^{2}+1} \right) \sqrt{\sqrt{{a}^{4}-{a}^{2}+1}-a} }{ 2 (a-1) \left( {a}^{2}-1 \right) \sqrt {{a}^{4}-{a}^{2}+1}}
	\equiv - t_3 \Delta\cI_{\pi} < 0 ,
	\label{disc_cI}
\end{equation}
where we introduced the notation $\Delta\cI_{\pi}$.
Taking this into account and since the eigenfunction should be continuous also at
$\theta = \pi$, we define
\begin{eqnarray}
	\cT_{3}^{(t_3)}(\theta, a) &\equiv& N(a) \exp \left\{
	i t_3 \left[\cI(\theta,a) - \cI(0,a) + \Theta \left(\theta - \pi \right) \Delta\cI_{\pi}  \right]
	+ \cR (\theta,a) - \cR(0,a)
	\right\} \nonumber \\
	&=& N(a) \frac{\sqrt{2}\left( 1 +a \right)^{3/2}}{\sqrt{\left( \cos\theta +a \right)
	\left| (3 \cos^2\theta+1) a + 2 \cos\theta ({a}^{2}+1) \right|}}
	e^{i t_3 \left[\cI(\theta,a) - \cI(0,a) + \Theta \left(\theta - \pi \right) \Delta\cI_{\pi}  \right]}
	, \label{eig_fcn}
\end{eqnarray}
where $N(a)$ is a normalization constant that we shall calculate in Section~\ref{sec_projections}, $\Theta(x)$ is the Heaviside step function, and we changed the notation $\Phi$ by $\cT_{3}^{(t_3)}$ to distinguish the eigenfunction corresponding to the eigenvalue $t_3$ from the other primitives.
This new expression is equivalent to~(\ref{primitive_sol}), but the solution is shifted upwards by $\Delta\cI_{\pi}$ in the $[\pi,2\pi]$ region in order to “connect” the discontinuity at $\pi$ (see Fig.~\ref{fig_eval}).
Further, from the periodicity condition $\cT_{3}^{(t_3)}(2\pi, a) = \cT_{3}^{(t_3)}(0, a)$ we obtain the quantization condition for $t_3$,
\begin{equation}
	t_3 \equiv \frac{2\pi n}{\Delta\cI_{\pi}}
	= - \frac {\hbar r n}{10 m_p}
	\frac{ 4 (a-1) \left( {a}^{2}-1 \right) \sqrt {{a}^{4}-{a}^{2}+1}}{ \left( {a}^{2}-3\,a+1-\sqrt {{a}^{4}-{a}^{2}+1} \right) \sqrt{\sqrt{{a}^{4}-{a}^{2}+1}-a} }
	\equiv \frac {\hbar r n}{10 m_p} t_3^{(0)}
	, \quad {\rm where} \quad n = {\rm integer} .
	\label{qc_t3}
\end{equation}
The dependence of $t_3^{(0)}$ on $a$ is shown in Fig.~\ref{fig_eval}.
The eigenfunctions~(\ref{eig_fcn}) may be multiplied by any function of $\phi$, since
$\hat{T}_3^{(\theta)}$ does not act on this coordinate.

Because of its singularities, the eigenfunctions $\cT_{3}^{(t_3)} (\theta,a)$ are not
square integrable on the interval $[0, 2\pi]$.
Nevertheless, we shall see in the next section that the Dirac brackets
$\cT_{3}^{(t_3)} (\theta,a)$ are well defined as kernels of integral functionals over the test functions $\Phi \in L^2_{cycl}([0, 2\pi])$.

\section{The eigendistributions of $\hat{T}_3^{(\theta)}$} \label{sec_projections}

In what follows, we shall ignore the $\phi$ dependence of the wavefunctions and retain only the $\theta$ dependence, that we want to study.
In a similar manner to the momentum when defined on a circle, it turns out that
$\hat{T}_3^{(\theta)}$ is self-adjoint when defined on the Sobolev space corresponding to our system~\cite{PhysicaA.2021.Dolineanu}.
In fact, an analogous proof can show that the $\theta$ projection of the momentum is also self-adjoint when defined on this domain, so we will formally introduce it in this section.

General Sobolev spaces~\cite{Mazya:book} have the underlying set defined as
\begin{equation}
	W^{k,p}(I) \coloneqq \{\psi\in L^p(I)\cap W^k(I)| \psi^{(\alpha)}\subset L^p(I),\,\forall |\alpha|\leq k\} ,
\end{equation}
where $\psi^{(\alpha)}$ is the $\alpha$ order derivative of $\psi$, $I\subset \R$ is
open, $k\in \mathbb{N}$, and $1\leq p\leq \infty$.
$W^k$ is the set of all locally integrable functions that also have weak derivatives
of order $\alpha$ for all $|\alpha|\leq k$.
By a locally-$L^p(I)$ function $\psi$, we refer to functions for which there exists a locally-$L^p(I)$ function $\rho$ such that $\int_I \psi \phi' dx = -\int_I \rho \phi dx$ for all $\phi \in C^\infty$ that vanish at the limits of integration.
We introduce the usual definitions
\begin{subequations}
	\begin{equation}
		H^k \coloneqq W^{k,2} ,
	\end{equation}
	\begin{equation}
		L^2_{cycl}([0,2\pi])\coloneqq\{\psi \in L^2([0,2\pi])|\psi(0)=\psi(2\pi)\} ,
	\end{equation}
	\begin{equation}
		H^1_{cycl}([0,2\pi]) \coloneqq \{\psi \in H^1([0,2\pi]) | \psi(0)=\psi(2\pi)\}\subset L^2_{cycl}([0,2\pi])
		\equiv \Omega ,
	\end{equation}
\end{subequations}
but we shall use $\Omega$ instead of $H^1_{cycl}([0,2\pi])$, for simplicity. In this context, the wavefunctions of our system are $\langle \theta|\Phi \rangle \equiv \Phi (\theta) \in \Omega$.
In~\cite{PhysicaA.2021.Dolineanu}, it was proven that
$\hat{T}_3^{(\theta)} : H^1_{cycl}([0,2\pi]) \rightarrow L^2_{cycl}([0,2\pi])$ is self-adjoint.

Following the example of the momentum operator~\cite{EurJPhys.26.287.2005.Madrid}, by taking the eigenfunctions of $\hat{T}_3^{(\theta)}$ as kernels, we construct the elements corresponding to the toroidal moment in the dual $\Omega^*$ of $\Omega$ by defining
\begin{eqnarray}
	\langle t_3,a | \Phi \rangle &=&
	\int_{0}^{2\pi} d\theta (R+r\cos\theta) {\cT_{3}^{(t_3)}}^* (\theta,a) \Phi(\theta) .
	\label{sc_prod_theta1}
\end{eqnarray}
Here, similarly to how the momentum operator is usually treated, we define the toroidal moment kets $|t_3, a \rangle$ (and analogously the bras) by having
\begin{equation}
	\cT_{3}^{(t_3)} (\theta,a) = \langle \theta | t_3,a\rangle
\end{equation}
such that $\hat{T}_3^{(\theta)} |t_3, a \rangle = t_3 |t_3, a \rangle$.
Notice that~(\ref{sc_prod_theta1}) can also be written as
\begin{equation}
	\langle t_3, a | \Phi \rangle \equiv
	\int_{0}^{2\pi} (R+r\cos\theta)
	\left\langle t_3,a \right|\theta\rangle \Phi(\theta) d\theta .
	\label{def_t3_Phi}
\end{equation}

We now proceed to show how the integral in~(\ref{sc_prod_theta1}) is to be computed. In the expression for $\cT_{3}^{(t_3)}$~(\ref{eig_fcn}) we denote
\begin{equation}
	y \equiv \cI(\theta,a) - \cI(0,a) - \Theta \left(\theta - \pi \right) \Delta\cI_{\pi} \equiv f_a(\theta) ,
	\quad {\rm so} \quad
	\theta \equiv f_a^{-1}(y) .
	\label{def_var_y}
\end{equation}
From Fig.~\ref{fig_Imprimitive}, we observe that $y$ takes values from $-\infty$ to $\infty$
and $f_a(\theta)$ may be split into three branches:
$f_{a}^{(1)}(\theta):[0,\theta_0^{(1)}) \to (-\infty, 0]$,  $f_{a}^{(2)}(\theta):(\theta_0^{(1)},
\theta_0^{(2)}) \to (-\infty,\infty)$, and $f_{a}^{(3)}(\theta):(\theta_0^{(2)},2\pi] \to
[\Delta \cI_{\pi},\infty)$.
Therefore, changing the variable from $\theta$ to $y$, the expression to be integrated becomes
\begin{eqnarray}
	\cE &\equiv& (R+r\cos\theta) {\cT_{3}^{(t_3)}}^* (\theta,a) \Phi(\theta)
	= N^*(a) (R+r\cos\theta(y)) \Phi(\theta(y)) \frac{\sqrt{2}\left( 1 +a \right)^{3/2} e^{-i t_3 y}}{\sqrt{\left( \cos\theta +a \right) \left| (3 \cos^2\theta+1) a + 2 \cos\theta ({a}^{2}+1) \right|}}
	\nonumber \\
	&& \times \frac{d f_a^{-1}}{dy}
	= N^*(a) (R+r\cos\theta(y)) \Phi(\theta(y)) \frac{\sqrt{2}\left( 1 +a \right)^{3/2} e^{-i t_3 y}}{\sqrt{\left( \cos\theta +a \right) \left| (3 \cos^2\theta+1) a + 2 \cos\theta ({a}^{2}+1) \right|}} \left(\frac{d \cI}{d\theta}\right)^{-1}_{\theta(y)}
	\nonumber \\
	&=& - N^*(a) r C_0 2\sqrt{2} \sqrt{a+\cos\theta} (a-1)^2 (a + 1)^{5/2} \sqrt {{a}^{4}-{a}^{2}+1} \frac{(3 \cos^2\theta+1) a + 2 \cos\theta ({a}^{2}+1)}{\sqrt{\left| (3 \cos^2\theta+1) a + 2 \cos\theta ({a}^{2}+1) \right|}}
	\Phi(\theta) e^{-i t_3 y} ,
	\label{expr_int}
\end{eqnarray}
where $\theta \equiv \theta(y)$.
Plugging~(\ref{expr_int}) into the integral of Eq.~(\ref{sc_prod_theta1}) we get
\begin{eqnarray}
	&& \langle t_3,a | \Phi \rangle = r C_0 N^*(a) 2\sqrt{2} (a-1)^2 (a + 1)^{5/2} \sqrt {{a}^{4}-{a}^{2}+1}
	\nonumber \\
	&& \times \left\{ \int_{-\infty}^{0} dy\sqrt{(a+\cos\theta) \left[ (3 \cos^2\theta+1) a + 2 \cos\theta ({a}^{2}+1) \right]}
	\Phi(\theta) e^{-i t_3 y}
	\right. \nonumber \\
	&& + \int_{-\infty}^{\infty} dy\sqrt{- (a+\cos\theta) \left[ (3 \cos^2\theta+1) a + 2 \cos\theta ({a}^{2}+1) \right]}
	\Phi(\theta) e^{-i t_3 y}
	\nonumber \\
	&& \left. + \int_{\Delta \cI_{\pi}}^{\infty} dy\sqrt{(a+\cos\theta) \left[ (3 \cos^2\theta+1) a + 2 \cos\theta ({a}^{2}+1) \right]}
	\Phi(\theta) e^{-i t_3 y}
	\right\} . \label{sc_prod_theta2}
\end{eqnarray}
We can make the integrals symmetric, by displacing $y$ and redefining the function
$\theta'(y') = \theta(y)$ on different segments:
\begin{subequations} \label{defs_D2}
\begin{eqnarray}
	&& \theta \in \D_1 \equiv [0,\theta_0^{(1)}), \quad C_1(\theta,a)<0, \quad y=y'\in (-\infty, 0), \label{def_D1} \\
	&& \theta \in \D_2 \equiv (\theta_0^{(1)},\theta_0^{(2)}), \quad C_1(\theta,a)> 0, \quad y-\frac{\Delta \cI_{\pi}}{2} = y' \in(-\infty, \infty), \label{def_D2} \\
	&& \theta \in \D_3 \equiv (\theta_0^{(2)}, 2\pi], \quad C_1(\theta,a)<0, \quad y- \Delta \cI_{\pi} = y' \in (0, \infty), \label{def_D3}
\end{eqnarray}
\end{subequations}
and write
\begin{eqnarray}
&& \langle t_3,a| \Phi \rangle
 = r C_0 N^*(a) 2\sqrt{2} (a-1)^2 (a + 1)^{5/2} \sqrt {{a}^{4}-{a}^{2}+1}
\nonumber \\
&& \times \left\{ \int_{-\infty}^{0} \left. dy\sqrt{(a+\cos\theta) \left[ (3 \cos^2\theta+1) a + 2 \cos\theta ({a}^{2}+1) \right]}
\Phi(\theta) e^{-i t_3 y} \right|_{\theta \in \D_1}
\right. \nonumber \\
&& + e^{-i t_3 \frac{\Delta \cI_{\pi}}{2}} \int_{-\infty}^{\infty} \left. dy\sqrt{- (a+\cos\theta') \left[ (3 \cos^2\theta'+1) a + 2 \cos\theta' ({a}^{2}+1) \right]}
\Phi(\theta') e^{-i t_3 y} \right|_{\theta \in \D_2}
\nonumber \\
&& \left. + e^{-i t_3 \Delta \cI_{\pi}} \int_{0}^{\infty} \left. dy\sqrt{(a+\cos\theta') \left[ (3 \cos^2\theta'+1) a + 2 \cos\theta' ({a}^{2}+1) \right]}
\Phi(\theta') e^{-i t_3 y} \right|_{\theta \in \D_3}
\right\} \nonumber \\
%
%
%
&& = r C_0 N^*(a) 2\sqrt{2} (a-1)^2 (a + 1)^{5/2} \sqrt {{a}^{4}-{a}^{2}+1}
\nonumber \\
&& \times \left\{ \int_{-\infty}^{0} \left. dy\sqrt{(a+\cos\theta) |C_1(\theta,a)|}
\left[\Phi(\theta) e^{-i t_3 y} + \Phi(2\pi-\theta) e^{i t_3 y}\right]  \right|_{\theta \in \D_1}
\right. \nonumber \\
&& + e^{-i t_3 \frac{\Delta \cI_{\pi}}{2}} \int_{-\infty}^{0} \left. dy\sqrt{ (a+\cos\theta') |C_1(\theta',a)|}
\left[ \Phi(\theta') e^{-i t_3 y} + \Phi(2\pi-\theta') e^{i t_3 y} \right] \right|_{\theta' \in \D_2}
. \label{sc_prod_theta3}
\end{eqnarray}

In the asymptotic regime, when $\theta \to \theta_0^{(1)}$, we retain only the dominant term
for $\cI$ from Eq.~(\ref{primitive_sol}) and obtain
%
\begin{subequations} \label{asympt_y_theta1}
\begin{equation}
	y \approx \frac{1}{C_0}
	\frac{\sqrt{\sqrt{{a}^{4}-{a}^{2}+1}+a} \left( {a}^{2}-3a+1+\sqrt {{a}^{4}-{a}^{2}+1} \right)}{4 (a-1) \left( {a}^{2}-1 \right) \sqrt {{a}^{4}-{a}^{2}+1}}
	\ln \left[ {\frac{(a-1)}{\sqrt {\sqrt {{a}^{4}-{a}^{2}+1}-a}}} \frac{|\theta - \theta_0|}{2\cos^2(\theta_0^{(1)}/2)} \right] ,
	\label{asympt_y1}
\end{equation}
which gives
\begin{equation}
|\theta - \theta_0| \approx \frac{2 \sqrt{\sqrt {{a}^{4}-{a}^{2}+1}-a} \cos^2(\theta_0^{(1)}/2) }{(a-1)}
\exp \left[ y C_0
\frac{4 (a-1) \left( {a}^{2}-1 \right) \sqrt {{a}^{4}-{a}^{2}+1}}{\sqrt{\sqrt{{a}^{4}-{a}^{2}+1}+a} \left( {a}^{2}-3a+1+\sqrt {{a}^{4}-{a}^{2}+1} \right)} \right] .
\label{asympt_theta1}
\end{equation}
\end{subequations}
On the other hand, in the same limit we have
\begin{subequations}\label{divergences_int}
\begin{equation}
	|C_1(\theta,a)|
	= \frac {2\sqrt {2}}{3a}\sqrt{ \left( -{a}^{4} + 4{a}^{2} -1 +\sqrt{{a}^{4}-{a}^{2}+1}({a}^{2}+1) \right) ({a}^{4}-{a}^{2}+1)} \, \left| \theta-\theta_0^{(1)} \right| ,
	\label{taylor_C1}
\end{equation}
which implies that, in Eq.~(\ref{sc_prod_theta3}),
\begin{equation}
	\left|C_1(\theta,a)\right|^{1/2} \propto \left| \theta-\theta_0^{(1,2)} \right|^{1/2} ,
	\quad {\rm when} \quad
	\theta \to \theta_0^{(1,2)} .
	\label{asympt_C1}
\end{equation}
Furthermore, $\Phi(\theta)$ belongs to $\Omega$ which implies that, if $\Phi(\theta)$
is continuous in some neighborhoods of $\theta_0^{(1)}$ and $\theta_0^{(2)}$, then there are
\begin{equation}
	M_\Phi > 0, \quad m_\Phi > -1/2, \quad \rm {such\ that} \quad
	|\Phi(\theta)| < M_\Phi \left| \theta - \theta_0^{(1,2)} \right|^{m_\Phi} .
	\label{div_Phi}
\end{equation}
\end{subequations}
From Eqs.~(\ref{sc_prod_theta3}) and (\ref{divergences_int}), we obtain that if $\Phi$ is continuous in some neighborhoods of $\theta_0^{(1)}$ and $\theta_0^{(2)}$, then there is $M'_\Phi > 0$, such that
\begin{equation}
	\sqrt{(a+\cos\theta) |C_1(\theta,a)|} |\Phi(\theta)| < M'_\Phi \left| \theta - \theta_0^{(1,2)} \right|^{1/2 + m_\Phi} \propto e^{(1/2+m_\Phi) r}
	\label{conf_int}
\end{equation}
($1/2+m_\Phi > 0$, according to~\ref{div_Phi}), where in the last relation we used
 Eq.~(\ref{asympt_theta1}).
This implies that the integrals in Eq.~(\ref{sc_prod_theta3}) converge in $\theta_0^{(1)}$
and $\theta_0^{(2)}$ and therefore in general, if $\Phi \in \Omega = H^1_{cycl}([0,2\pi])$.

The considerations for the right-functionals, the elements of the space of
antilinear functionals over $\Omega$ (denoted $\Omega^\times$), are to
be defined in complete analogy to the left-functionals we have presented above.

We now have to normalize these distributions and prescribe how to move into the toroidal moment representation. The most commonly encountered unbounded operators, position and momentum in $L^2(\mathbb{R})$,
have a continuous spectra, and the changing from one representation to the other is done via
Fourier transforms, which are delta normalized~\cite{EurJPhys.26.287.2005.Madrid}.
However, as we have seen in Section~\ref{sec_eigfs}, the spectrum of $\hat{T}_3^{(\theta)}$ is quite peculiar because it is discrete. This means that, in order to apply a similar procedure as in~\cite{EurJPhys.26.287.2005.Madrid}, we need to write the resolution of identity as a sum when working in the toroidal moment space, despite having it as an integral in position space.

We proceed by considering the following integral between two kernels:
\begin{eqnarray}
&& \langle t_3, a| t_3', a\rangle \equiv \int_{0}^{2\pi} \langle t_3,a |\theta \rangle
\langle \theta | t_3, a \rangle d\theta
= r C_0 |N(a)|^2 4 (a-1)^2 (a + 1)^4 \sqrt {{a}^{4}-{a}^{2}+1}
\nonumber \\
&& \times \left\{ \int_{-\infty}^{0} \left. dy e^{i (t_3' - t_3) y} \right|_{\theta \in \D_1}
+ e^{i (t_3'-t_3) \frac{\Delta \cI_{\pi}}{2}} \int_{-\infty}^{\infty} \left. e^{i (t_3' - t_3) y} \right|_{\theta \in \D_2}
+ e^{i (t_3'-t_3) \Delta \cI_{\pi}} \int_{0}^{\infty} \left. dy e^{i (t_3' - t_3) y} \right|_{\theta \in \D_3}
\right\} \nonumber \\
&& = r C_0 |N(a)|^2 4 (a-1)^2 (a + 1)^4 \sqrt {{a}^{4}-{a}^{2}+1} \left[1 + e^{i (t_3' - t_3) \frac{\Delta \cI_{\pi}}{2}} \right] \int_{-\infty}^{\infty} dy e^{i (t_3' - t_3) y}
. \label{sc_prod_theta_T3}
\end{eqnarray}
Obviously, the integral is infinite for $t_3'=t_3$ and does not converge if $t_3' \ne t_3$.
Therefore, we may define the previous bracket as
\begin{subequations} \label{defs_prod_lim}
\begin{eqnarray}
&& \left\langle t_3,a \right| \left. t_3',a \right\rangle_{lim}
\equiv r C_0 |N(a)|^2 4 (a-1)^2 (a + 1)^4 \sqrt {{a}^{4}-{a}^{2}+1} \left[1 + e^{i (t_3' - t_3) \frac{\Delta \cI_{\pi}}{2}} \right]
\nonumber \\
&& \times \lim_{y_{max} \to \infty} \frac{1}{2y_{max}} \int_{-y_{max}}^{y_{max}} dy e^{i (t_3' - t_3) y}
=
\begin{cases}
2 r C_0 |N(a)|^2 4 (a-1)^2 (a + 1)^4 \sqrt {{a}^{4}-{a}^{2}+1} , & {\rm if} \quad t_3'=t_3 , \\
0 , & {\rm if} \quad t_3'\ne t_3 .
\end{cases}
\label{sc_prod_theta_T3_lim}
\end{eqnarray}
Then, from the condition $\left\langle t_3,a \right|\left. t_3',a \right\rangle_{lim} = \delta_{t_3 t_3'}$ we define the normalization constant,
\begin{equation}
	|N(a)|^2 \equiv \frac{1}{8 r C_0 (a-1)^2 (a + 1)^4 \sqrt {{a}^{4}-{a}^{2}+1}} .
	\label{norm_Na}
\end{equation}
\end{subequations}
This allows us to normalize the distributions in~(\ref{sc_prod_theta1}) by redefining the kernels $\cT_{3}^{(t_3)} (\theta,a)$ as multiplied by the normalization factor~(\ref{norm_Na}).

With this setup, we can now explicitly write how any wavefunction $\Phi$ may be transformed from the position representation, $\langle \theta|\Phi \rangle \equiv \Phi (\theta) \in \Omega$, into the $\hat{T}_3^{(\theta)}$ representation:
\begin{subequations} \label{transf_L2_V_tot}
\begin{equation}
	\left| \cT_3^{(\Phi)} \right\rangle \equiv \sum_{t_3} \left| t_3, a \right\rangle \left\langle t_3, a | \Phi \right\rangle .
	\label{transf_L2_V}
\end{equation}
In this representation, $\hat{T}_3^{(\theta)}$ is equivalent to a multiplication-by-number operator,
\begin{equation}
	\hat{T}_3^{(\theta)} \left| \cT_3^{(\Phi)} \right\rangle = \sum_{t_3} t_3 \left| t_3, a \right\rangle \left\langle t_3, a | \Phi \right\rangle .
	\label{T3_transf_L2_V}
\end{equation}
\end{subequations}

\section{Conclusions} \label{sec_conclusions}

We found analytical expressions for the eigenfunctions and eigenvalues of the projection
$\hat{T}_3^{(\theta)}$ of the toroidal dipole operator of a particle in a thin layer bent
in the form of a torus (layer thickness is much smaller than the minor radius of the torus,
whereas the major radius is bigger than the minor radius).
From these expressions, we found the quantization relations for the eigenvalues $t_3$,
which are the measurable values of $\hat{T}_3^{(\theta)}$.

The eigenfunctions are not integrable in modulus square and therefore they represent
distributions over the space $H^1_{cycl}([0,2\pi])$.
We find a normalization criterion for these eigendistributions.
Having this setup, we prescribe the changing of the wavefunctions from the position
representation to the $\hat{T}_3^{(\theta)}$ representation.

\section{Acknowledgments} \label{sec_ack}

This work has been financially supported by the ELI-RO contract 81-44 /2020.
Travel support was also provided by the Romania-JINR collaboration projects positions 19, 22, Order 366/11.05.2021.
Discussions with Dr. Cristinel Stoica are gratefully acknowledged.


\end{document}